\documentclass[preprint,
 amsmath,amssymb,
 aps,
]{revtex4-1}

\usepackage{graphicx}
\usepackage{dcolumn}
\usepackage{bm}

\begin{document}

\title{An accurate and efficient Lagrangian sub-grid model}

\author{Irene M. Mazzitelli} \affiliation{Dip. Ingegneria
  dell'Innovazione, Univ. Salento, 73100 Lecce, Italy \& CNR-ISAC, Via Fosso del Cavaliere 100, 00133 Roma, Italy}
\author{Federico Toschi} \affiliation{Department of Applied Physics
  and Department of Mathematics \& Computer Science, Eindhoven
  University of Technology, Eindhoven, 5600MB, The Netherlands\\and
  CNR-IAC, Via dei Taurini 19, 00185 Rome, Italy}
\author{Alessandra S. Lanotte} \thanks{Corresponding author: a.lanotte@isac.cnr.it}
\affiliation{CNR ISAC and INFN, Sez. di
  Lecce, Str. Prov. Lecce Monteroni, 73100 Lecce, Italy }

\date{\today}
\begin{abstract}
A computationally efficient model is introduced to account for the
sub-grid scale velocities of tracer particles dispersed in
statistically homogeneous and isotropic turbulent flows. The model
embeds the multi-scale nature of turbulent temporal and spatial
correlations, that are essential to reproduce multi-particle
dispersion. It is capable to describe the Lagrangian diffusion and
dispersion of temporally and spatially correlated clouds of
particles. Although the model neglects intermittent corrections, we
show that pair and tetrad dispersion results nicely compare with
Direct Numerical Simulations of statistically isotropic and
homogeneous $3D$ turbulence. This is in agreement with recent
observations that deviations from self-similar pair dispersion
statistics are rare events.
\end{abstract}

\pacs{}

\maketitle

\section{\label{sec:intro}Introduction}
The transport of particles in turbulent flows is strongly sensitive to
the multi-scale and multi-time fluctuactions of the turbulent Eulerian
velocity. For this reason, the dispersion of particles poses
extraordinary challenges when the complexity of the flow geometry or
the large Reynolds numbers requires the use of turbulence models. In
particular, the modelisation of the small Eulerian scales can
significantly alter the dynamics of particle dispersion. Particle
dispersion in turbulence, either from extended or from localized
sources \cite{Silv2010,SBTPRL2012}, is a very common phenomenon of
practical importance for atmospheric as well as for many applied
problems \cite{toschi_bode}. Among the many possible examples, we
remind here the dynamics and the spatial distribution of pollutants
and pollen in the atmosphere \cite{MH2006,CM2011,LMR2008,ML2012} or
oceanic flows \cite{berti2011,Palatella2013}, the formation and the
dynamics of small rain droplets in clouds \cite{shaw2003}, the
dynamics of colloidal aggregates in turbulence \cite{BMB2008,BBL2012},
the combustion of fuel droplets and the formation of soot particles in
engines \cite{PA2002}.

The development of turbulence models and closures, to describe the
effect of the {\it unresolved} or sub-grid scale (SGS) features of the
Eulerian vector or scalar fields, has a long history dating back to
Lilly and Smagorinsky (see \cite{segaut}). It is fair to say that
nowadays there are a number of well-established {\it classical} SGS
models for homogeneous and isotropic turbulence (HIT), adapted and
extensively tested under a variety of conditions \cite{MK2000}, as
well as more recent proposals keeping into account the phenomenology
of turbulence beyond HIT (see e.g.,
\cite{MoengSullivan,ToschiLeve}).\\ The development of sub-grid models
for Lagrangian turbulence has a relatively shorter history, partly due
to the lack of accurate experimental and direct-numerical simulation
measurements of Lagrangian statistics in high-Reynolds number
flows. The recent availability of a large amount of Lagrangian
statistics measurements in HIT
\cite{Ott,Xunature,PintonPRL,IK2002,YeBo2004,BoffSok3D,PRL2004,Pof2005}
has allowed to quantitatively establish the phenomenological picture
for tracers (reviewed in \cite{Yeungrev,Sawfordrev,Collinsrev}), and
partially also for inertial point-particles \cite{JFM2010}. The
knowledge borrowed from experiments and direct numerical simulations
has then promoted new research on Lagrangian sub-grid scale models for
tracers, and inertial particles also (see e.g.,
\cite{Weil2004,LMR2008,JH2013}). The effects of small-scale temporal
and spatial correlations on the dynamics of particles are an important
problem \cite{FGVrev}. In particular beyond {\it classical}
measurements of the Lagrangian dynamics of a single particle and of
particle pairs, the geometric features of multiparticles dispersion
has also been investigated
\cite{CPS99,Pof2005multi,Xu2007,HYS2011}.\\ Within the complex picture
of Lagrangian dynamics, one of the most important point is that
Lagrangian turbulence is more intermittent than Eulerian turbulence
\cite{PRLall} and, as a result, one may have to pay additional care
when using Gaussian models to model the Lagrangian velocity
fields. Moreover, in HIT, the relative dispersion of tracers is mainly
dominated by small-scale fluid motions: if these are neglected,
particle pairs disperse at a much slower rate than the actual one
(ballistic vs Richardson dispersion).

Traditionally, Lagrangian SGS motions are described by means of
stochastic models. These are based on stochastic differential
equations for the evolution of the velocity, assumed to be Markovian,
along a particle trajectory. These can be built up for single particle
trajectories \cite{Tho87}, two-particle \cite{KD88,Tho90,K97} and
four-particle dispersion \cite{DevTho2013,Dev2013}. The literature on
the topic is vast and we cannot review it here. What is important for
the present discussion is that stochastic models for two-particle
dispersion are generally inconsistent with single particle statistics,
so that depending on the problem at hand one has to change model.\\ A
different approach was developed in Lacorata et al. \cite{LMR2008},
where a multiscale kinematic velocity field was introduced to model
turbulent relative dispersion at sub-grid scales. The authors
exploited Lagrangian chaotic mixing generated by a nonlinear
deterministic function, periodic in space and time. This approach
differs from kinematic models (e.g.  \cite{Fung1992,FV1998}), as it
reproduces the effect of large-scale sweeping on particle trajectories
\cite{Chaves,DevThom2005,Eyinksweep}.\\ In the context of wall-bounded
flows, Lagrangian SGS schemes have been proposed in terms of
approximate deconvolution models based on the Eulerian field (see
e.g., \cite{K2006}), or in terms of force-based models
\cite{soldati_salvetti}. Observables capable to discriminate between
model error and drift induced errors were proposed
\cite{toschi_donini}. \\Most of these models rely on the knowledge of
the resolved Eulerian velocity field, however we note that models to
solve the Lagrangian dynamics self-consistently without an underlying
Eulerian velocity field have been proposed. This is for example the
idea behind Smoothed Particle Hydrodynamics (SPH), i.e. a purely
Lagrangian scheme to solve the Navier-Stokes equation, recently
reviewed in \cite{Mon2012}. In Smoothed Particle Hydrodynamics,
instead of solving the fluid equations on a grid, it is used a set of
particles, whose equations of motion are determined from the continuum
Navier-Stokes equations.

An important issue concerns the possibility to build up models
accounting for multi-particle dispersion, ${\cal N}>2$, going beyond
the pair separation dynamics. Multi-particle Lagrangian models
invariably need to incorporate a mechanism correlating the
sub-grid-scale velocities of the particles. Different approaches are
possible. In Sawford {\em et al.}~\cite{Sawford2013}, a two-particle
stochastic model for $3D$ Gaussian turbulence \cite{Tho90,bor94} has
been generalised to the problem of $n$ tracers: these are constrained
by pair-wise spatial correlations, implying that multi-point
correlations are neglected. Interestingly, the model shows a good
agreement of multi-point statistics with direct numerical simulations
results. Alternatively, Burgener {\em et al.} \cite{Burgener2012}
proposed to build spatial correlations between the fluid particles by
minimasing of a Heisenberg-like Hamiltonian. In the Hamiltonian, the
two-object coupling function is distance dependent and with a power
law behaviour. Ballistic separation and Taylor diffusion regimes in
pair dispersion are clearly observed, while turbulent inertial-range
dispersion {\it \'a la} Richardson is observed in specific conditions.

In this manuscript we introduce a novel, accurate and computationally
efficient Lagrangian Sub-Grid Scale model (LSGS) for the dispersion of
an arbitrary number of tracers in $3D$ statistically homogeneous and
isotropic turbulent flows. The model is purely Lagrangian and it
defines and evolves the velocities of tracers at their positions. The
trajectories of ${\cal N}$ particles are simply obtained by
time-integrating the Lagrangian velocities. It is primarily meant to
reproduce Lagrangian dispersion as sub-grid scales, but it may be used
as well as a rudimental Lagrangian Navier-Stokes solver, much in the spirit of
SPH.\\ The model encodes velocity fluctuations that scale in space and
in time consistently with Kolmogorov 1941 \cite{frisch}, hence without
intermittentcy corrections, and is self-consistent for an arbitrary
number or density of tracers. An essential prescription for the model
is the capability to correctly reproduce single-particle absolute
diffusion together with multi-particle dispersion. For the latter, we
require proper reproduction of inertial range pair dispersion
(Richardson dispersion \cite{Sawfordrev,Pof2005,Collinsrev}), as well
as dynamics and deformation of tetrads.\\ In a nutshell, the idea of
our LSGS model is to define a multiscale relative velocity difference
between two tracers, consistent with Kolmogorov inertial range
scaling. Such velocity difference, characterized by the proper eddy
turnover time, is able to reproduce Richardson dispersion for a single
pair of tracers. The model is built up in a similar spirit of what
done in \cite{LMR2008} for tracer pair dispersion, but it is capable
of ensuring consistent correlations between an arbitrary number of
tracers according to their positions and relative distances. Space
correlations ensure that tracers close in space will experience very
similar SGS velocities. Beyond pair dispersion, we also quantitatively
validate the temporal evolution and dispersion properties of groups of
four particles (tetrads), against Direct Numerical Simulations results
\cite{Pof2005multi}.

The paper is organized as follows. In Section~\ref{sec:model}, we
introduce the LSGS model for an arbitrary number of tracers, and with
an arbitrary large inertial range of scales. In
Sec.~\ref{sec:results}, we specify the model parameters and discuss the
results for absolute, pair and tetrad dispersion. Last section is
devoted to concluding remarks.

\section{The Lagrangian Particle Model}
\label{sec:model}
In large-eddy simulations, the full tracer velocity is defined as the
sum of the resolved Lagrangian velocity component, ${\bm V}_i({\bm
  x}_i, t)$, and the sub-grid-scale contribution, ${\bm v}_i({\bm
  x}_i(t),t)$. The larger scale components of the velocity,
characterized by larger correlation times, sweep the smaller ones thus
advecting both particles and small-scale eddies. This is a crucial
feature of Lagrangian turbulence, sometimes neglected in synthetic
models of Eulerian turbulence, that incorrectly describe pair
dispersion \cite{Chaves,DevThom2005,Eyinksweep}.\\ The Lagrangian
sub-grid-scale model describes the $3D$ velocity, ${\bm v}_i({\bm
  x}_i(t),t)$, at the position, ${\bm x}_i(t)$, of the $i-${\em th} of
the ${\cal N}$ tracer particles. The velocity fluctuations along each
particle trajectory are the superposition of different contributions
from eddies of different sizes. These eddies constitute a turbulent
field, decomposed for convenience in terms of logarithmically spaced
shells.\\ We consider the velocity of a tracer built as the sum of a
set of fluctuations, ${\bm u}_n$, of index $n$ associated to a
equispaced set of lengthscales, $l_n$. Given the largest length scale
of the flow, $L_0$, smaller scales are defined as:
\begin{equation}
l_n = \frac{L_0}{\lambda^n}, \,\quad\quad n=0,\dots,N_m-1\,,
\label{eq:scale}
\end{equation}
where $N_m$ is the total number of modes, scales are logaritmically
equispaced and the factor $\lambda > 1$ is conventionally chosen as
$\lambda = 2^{1/4}$. Length-scales correspond to wave-numbers
$k_n\,=\, 2 \pi /l_n$, so that the velocity amplitudes and the
associated turn-over times are defined as 
\begin{equation}
u_n = q_0 \, k_n^{-1/3},\qquad \tau_n = l_n /u_n\,,
\label{eq:defK41}
\end{equation}
where $q_0$ is associated with the amplitude of the large-scale
velocity.

\subsection{Implementation of the LSGS}
\label{sec:modeldetails}
With the aim of making the description as clear as possible, we
consider that the model is best illustrated by the following two-steps
procedure:\\ {\em Step 1}: At time $t$ the positions ${\bf x}_i(t)$ of
all ${\cal N}$ particles are given. The algorithm then generates -for
each tracer $i$ and for each lengthscale $l_n$- a {\it first} set of
velocity vectors $\boldsymbol{\zeta}_n^{(i)}(t)$ (the three velocity
components along the space directions $x,y,z$ that are chosen
independently from each other). Each of these velocities is the
outcome of an Ornstein-Uhlenbeck (OU) process with correlation time
$\tau_n = l_n/u_n$ and variance $u_n^2$, at the scale $l_n$. The time
evolution of the OU process, for each spatial component of the
velocity field of the $i-${\em th} particle at lengthscale $l_n$, is
obtained according to \cite{gil96}\,
\begin{equation}
\zeta_n^{(i)}(t+dt)\, =\, \zeta_n^{(i)}(t)\, e^{-dt/\tau_n} +
u_n\sqrt{1\,-\,e^{-2dt/\tau_n}}\, g\,,
\label{eq:ou}
\end{equation}
where, for simplicity, we have omitted the sub-script for the spatial
components. In (\ref{eq:ou}), the variable $g$ is a
random number, normally distributed in the range $[0;1]$. According to
our definitions, each OU process is a normally distributed, random
variable, $\zeta_n^{(i)}(t)$, whose mean value $\mu_n^{(i)}$ and
standard deviation $\sigma_n^{(i)}$ are\,:
\begin{eqnarray}
\mu_n^{(i)} &=& \zeta_{n_0}\,\exp{\left[-(t-t_0)/\tau_n
    \right]}\,\\ \sigma_n^{(i)} &=& u_n\,
\sqrt{1-\exp{\left[-2(t-t_0)/\tau_n\right]}}.
\label{eq:oumeanvar}
\end{eqnarray}
The equilibrium time $O(\tau_0)$ is needed for each mode to relax to a
zero mean velocity and to the variance $u_n^2$. The velocity
associated to the $i-${\em th} tracer particle is the superposition of
$N_m$ modes given by
\begin{equation}
\label{eq:zeta}
 {\bf v}^{(i)}(t) = \sum_{n=0}^{N_m-1}
 \boldsymbol{\zeta}_n^{(i)}(t)\,.
\end{equation}
So doing, each particle has a multiscale, single-point velocity field
which has the physical time correlations, but which does not respect
space correlations yet. Indeed, based on the above algorithm, very
different velocity fields could be assigned to particles residing in
very close spatial position.
\begin{figure}[!t]
\centerline{\includegraphics[width=10cm]{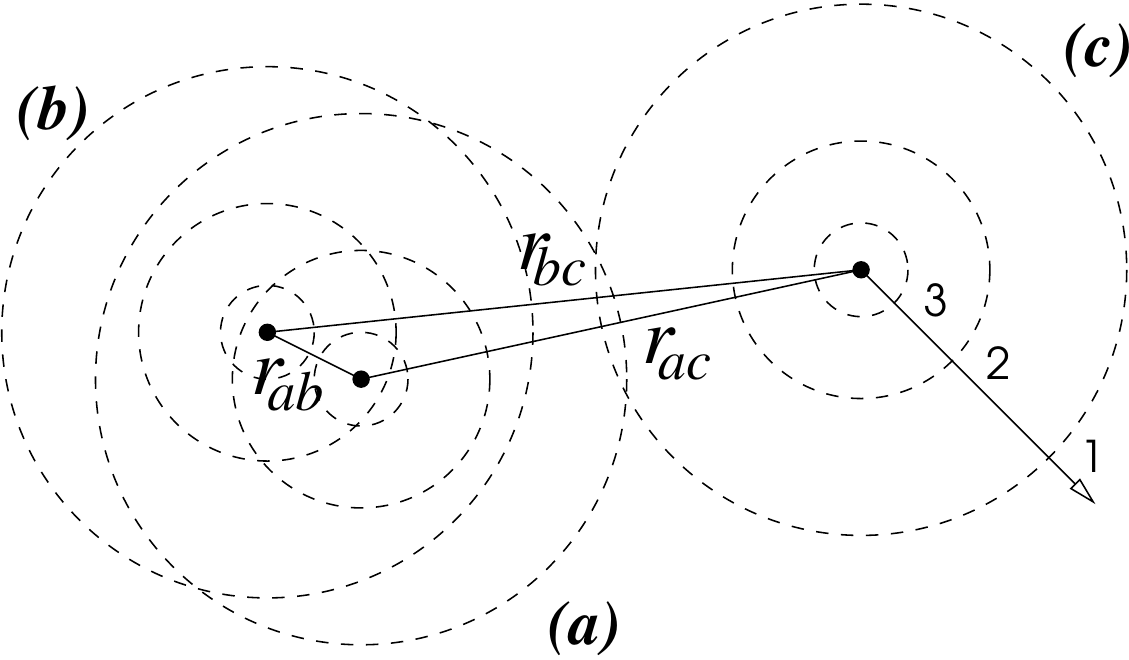}}
\caption{A sketch of the model. The sketch is for the case of three
  particles (${\bf a}$, ${\bf b}$ and ${\bf c}$) with velocity fields
  composed by three eddies ($1$, $2$ and $3$, that corresponds to the
  three circles of diameter, respectively, $l_1$, $l_2$ and
  $l_3$). Tracers positions are indicated by black dots, while dashed
  circles are {\it correlation radii} relative to the eddies with
  diameter $l_1$, $l_2$ and $l_3$ (corresponding to shells 1,2, and
  3). The largest eddy described by the sub-grid-scale model may
  correspond to the smallest resolved eddy of the Eulerian Large-eddy
  simulation. In case of a purely Lagrangian evolution, i.e. without a
  large-scale model or a LES, the largest eddy corresponds to the
  integral scales of the system, $L_0$.  The three modes of the
  velocity of particle ``${\bf a}$'' are computed by modulating the OU
  processes associated to particle ``${\bf a}$'', -
  $\boldsymbol{\zeta}_i^{({\bf a})}$ with $i=1,2,3$ -, with the OU
  processes attached to particles ``${\bf b}$'' and ``${\bf
    c}$''. This is done, in this simple example, according to: $ {\bf
    v}_1^{({\bf a})} = \boldsymbol{\zeta}_1^{({\bf
      a})}+\boldsymbol{\zeta}_1^{({\bf b})}(1-{r_{ab} \over
    l_1})+\boldsymbol{\zeta}_1^{({\bf c})}(1-{r_{ac} \over l_1})$; $
  {\bf v}_2^{({\bf a})} = \boldsymbol{\zeta}_2^{({\bf
      a})}+\boldsymbol{\zeta}_2^{({\bf b})}(1-{r_{ab} \over l_2})$; $
  {\bf v}_3^{({\bf a})} = \boldsymbol{\zeta}_3^{({\bf a})}$, where we
  have kept into account the fact that (a) and (b) overlap with shells
  1 and 2, while (a) and (c) overlap only with shell 1.}
\label{fig:sketch}
\end{figure}

{\em Step 2}: In order to build up the proper spatial correlations and
establish a correspondence between the modelled particle velocities
and the two-point Eulerian statistics, we redefine for the $i-th$
particle the fluctuation associated to the $n$-th mode as follows\,,
\begin{equation}
\label{eq:newzeta}
{\bf v}_n^{(i)}(t) = \sum_{j=1}^{{\cal N}}
\boldsymbol{\zeta}_n^{(j)}(t) \cdot  
\left( 1 - f_{l_n}(|{\bf x}_i - {\bf x}_j|) \right)\,.
\end{equation}
Here the decorrelation function is such that $f(r)\propto r$ for $r
\ll 1$ and $ f(r) \simeq 1$ for $r \gg 1$. Note that in
(\ref{eq:newzeta}), the $n-th$ mode velocity fluctuation for the
$i-$th particle is {\it determined} by the value of the $n-th$ mode
velocity fluctuation of the particle $j$, with $j=1, \dots, {\cal N}$
spanning over the entire particle ensemble. Clearly, only particles
closeby matter, while particles located very far from $i$ will not
matter.

The particle velocity resulting from the contributions of different
eddies is then evaluated as:
\begin{equation}
{\bf v}^{(i)}(t) = \sum_{n=0}^{N_m-1}\frac{1}{{\cal A}_n^{(i)}} {\bf v}_n^{(i)}(t).
\label{eq:newvel}
\end{equation}
The normalization factor that preserves the variance is given by
\begin{equation}
{\cal A}_n^{(i)} = \sqrt{\sum_{j=1}^{\cal N} 
(1 - f_{l_n}(|{\bf x}_i - {\bf x}_j|))^2}.
\label{eq:norma}
\end{equation}
As a result of this procedure, space correlations are introduced among
the velocities of the whole particle ensemble. Particle velocities are
no longer statistically independent, but behave as {\it responding} to
the same local eddy fluctuations. In Figure \ref{fig:sketch}, a graphic
sketch of the model is given for the case of three Lagrangian tracers,
with a few modes velocity.

Note that, from (\ref{eq:zeta}), the variance of each velocity
component along $x,y,z$ is:
\begin{equation}
\langle {v'}^{(i)}(t)^2 \rangle = \sum_{n=0}^{N_m-1} \langle {\bf
  \zeta}_n^{(i)}(t)^2\rangle\,,
\label{eq:totvariance}
\end{equation}
since modes are independent, $\langle {\bf \zeta}_n^{(i)}(t){\bf
  \zeta}_{n'}^{(i)}(t) \rangle = 0$ for $ n \neq n'$. If by physical
arguments, we require a velocity field with root-mean-square values to
$v_{rms}=u_0$, it is enough to introduce in the velocity definition
given in equation (\ref{eq:zeta}) or equation(\ref{eq:newvel}), the
norm ${\cal F}$ defined as\,,
\begin{equation}
{\cal F}^2 = \frac{u_0^2}{\sum_{n=0}^{N_m-1} u_n^2}\,.
\label{eq:totnorm}
\end{equation}
It is worth mentioning that the model has a tunable free parameter,
corresponding to the turbulent energy dissipation rate. If used as a
sub-grid, such parameter is provided by the large-scale model (e.g., a
large-eddy simulation), and can then be used to fix the ratio among
large-scale velocity and length values. In the absence of a
large-scale model, an energy dissipation rate can be fixed up to
constant $O(1)$.

\begin{figure}[!t]
\centerline{\includegraphics[width=10cm]{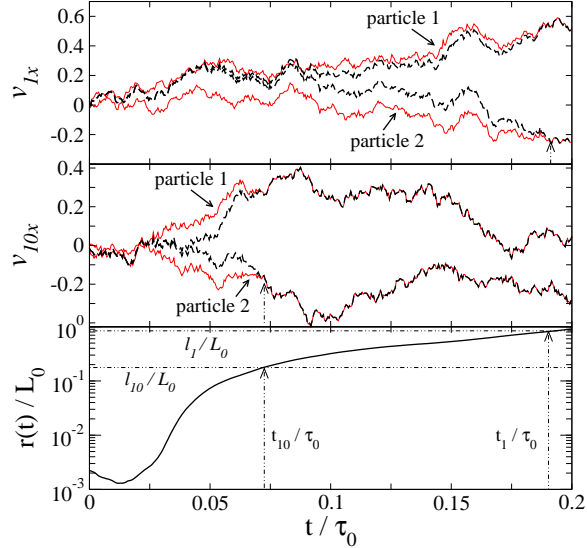}}
\caption{Particle velocity modulation in the case of a single pair,
  initially placed at a separation distance $r_0$ smaller than the
  smallest eddy of the fluctuating velocity $r_0 < l_{N_m-1}$, with
  $N_m=31$. Top: velocity fluctuation ($x$- component only) associated
  to the mode $n=1$. Middle: the same but for the mode $n=10$. In both
  cases, the curves represent: particle $1$ OU process (red continuous
  line), particle $2$ OU process (red continuous line), particle $1$
  and $2$ modulated velocity (black dashed lines). Bottom: the time
  evolution of the two particles separation ${\bf r}(t)= {\bf x}_1(t)
  - {\bf x}_2(t)$.}
\label{fig:modes}
\end{figure}

We note that from a computational point of view, the simplest
implementeation of the LSGS model scales as ${\cal N}^2$, but it can
be easily optimized with standard Molecular Dynamics algorithms
(e.g. by using a linked list).\\ To validate the model against
observations for statistically homogeneous and isotropic $3D$
Lagrangian turbulence, we consider in the following the most
challenging case corresponding to ${\bm V}=0$, when the sub-grid-scale
model is solely responsible for the dynamics of tracers at all
scales. Thus, to simplify the text, from now on the term ``sub-grid''
is dropped and we speak of tracers velocities; moreover we adopt the
shorthand notation ${\bm v}_i(t) \equiv {\bm v}_i({\bm x}_i(t),t)$.\\A
first glance on the behaviour of the model can be appreciated in
Figure~\ref{fig:modes}. Here we consider the results of a simulation
with only a pair of particles, with relative distance $r_0$ smaller
than the smallest eddy $l_{N_m-1}$ of the velocity field. Simulation
parameters are summarized in Table \ref{table1}, case $N_m= 31$. For
plotting purpose, we selected two modes, namely mode $n=1$ of length
scale $l_1=L_0/\lambda$ and mode $n=10$, with
$l_{10}=L_0/(\lambda)^{10}$. The OU processes, $\zeta_{nx}^{(i)}(t)$
given by (\ref{eq:ou}), resulting from the first step of the procedure
are compared with their modulation due to nearby particle, see
equation (\ref{eq:newzeta}). In the beginning, when particles are very
close, they possess the same velocity (in the figure, the $x$
component only is shown). When their separation becomes larger than
the mode of length-scale $l_n$ (in the example $n=1$ and $n=10$)
particle velocities decorrelate, and thus each particle velocity
collapses on its single-particle behaviour. As expected, decorrelation
is faster for the $n=10$ with respect to $n=1$ mode, since for the
former the eddy turnover time is smaller than for the latter.

Before discussing the performances of the model in the case of
particle diffusion and dispersion, a more general remark is needed. As
it is well know, the satisfaction of the fluid governing equation
yields to constraints on the particle system. For instance, the mean
continuity equation $ \nabla \cdot {\bf u}({\bf x},t)= 0$ is satisfied
when the particle density in space is uniform and constant at all
points \cite{pope00}. In stochastic approaches to Lagrangian particle
velocity, physical information is used to constrain the form of the
equation. The noise term has to be diffusive, while the drift term can
be specified on the basis of the Eulerian statistics of the flow. The
physical request is that an initially uniform particle distribution
will remain such, after Lagrangian evolution (from the Eulerian point
of view, a well-mixed scalar field remains so). In $3D$ there is no
unique form for the drift term, but there are a number of available
solutions \cite{Tho90,bor94}.\\Tracer uniform distribution is clearly
a crucial feature for a SGS model for incompressible turbulence. In
the Appendix, we discuss a series of test we performed to assess the
spatial distribution properties of the Lagrangian tracers, or in other
words to assess the {\it incompressibility} of the particle velocity
field.

\section{Results}
\label{sec:results}
\begin{figure}
\centerline{\includegraphics[width=10cm]{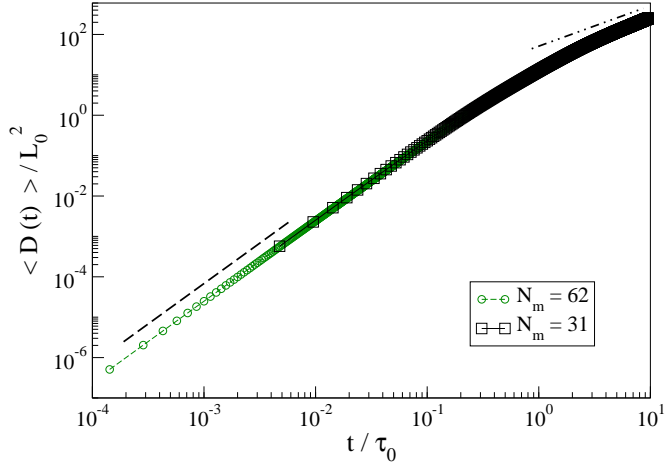}}
\caption{Log-log plot of tracers absolute dispersion for the simulations
  with $N_m=31$ (black squares) and $N_m=62$ (green
  circles). Simulation with $N_m=31$ modes was carried on for a longer
  time, whereas simulation with $N_m=62$ extendes to smaller times,
  owing to the smaller time step. The straight lines indicate,
  respectively, the $t^2$ ballistic behaviour (dashed line), and the
  diffusive $t$ behaviour (dashed dotted line).}
\label{fig:absdisp}
\end{figure}
We now discuss the results of two sets of numerical simulations,
characterized by different values of the total number of modes $N_m$,
at fixed values of the integral scale $L_0$ and root-mean-square
velocity $u_0$. Increasing the number of modes at fixed $L_0$ and
$u_0$ results in an extension of the inertial range of turbulence. For
each set of numerical simulations, mean values are computed by
ensemble averaging over $50$ simulations, each containing $100$
particle pairs. Particles are initially uniformly distributed in
space, and such that the initial pair separation is smaller than the
smallest eddy in the velocity, $l_{N_m-1}$. Their total number is
${\cal N} = 10^4$. The simulation parameters are reported in Table
\ref{table1}. Our model has the inertial range correlation of $3D$
turbulence built in, and thus cannot reproduce dissipative range
behaviours where particle dispersion is exponential and the rate of
separation is the leading Lyapunov exponent \cite{Pof2005}. We note
that it is however possible, when needed, to modify the model and
include a viscous range of scales: a way to do it is to use a Fourier
implementation of the so-called Batchelor-like parametrization of the
fluid velocity \cite{Men1996}.

We first consider the absolute dispersion, that is the mean displacement
of a single particle with respect to its initial position. The
statistical behaviour is expected to be ballistic for correlated
scales, followed by simple diffusion {\it \'a} la Taylor at scales
larger than the velocity integral scale $L_0$. To this aim we
compute:
\begin{equation}
D(t) = \langle \left[ {\bf x}(t) - {\bf x}(0)\right]^2 \rangle,
\label{ed:diffusion}
\end{equation}
where ${\bf x}(t)$ is the position at time $t$ of a particle that was
in ${\bf x}(0)$ at the initial time $t=0$ and the brackets $\langle
\cdot \rangle$ indicate ensemble average. From
Figure~\ref{fig:absdisp}, we observe that single particle diffusion is
well represented by the model. At time scales of the inertial range,
the slope is well approximated by the ballistic $t^2$ power law,
whereas at large times, $t > \tau_0$, the diffusive scaling law sets
in.
\begin{table}
\begin{tabular}{cccccc} 
 $N_m$ &$q_0$ & $L_0$ & $l_{N_m-1}$ & $\tau_0$ & $\tau_{N_m-1}$\\
\hline
$31$ &$0.4$ & $10$ & $5.5 \cdot 10^{-2}$ & $21$ & $6.7 \cdot 10^{-1}$ \\
$62$ &$0.4$ & $10$ & $2.6 \cdot 10^{-4}$ & $21$ & $1.9 \cdot 10^{-2}$ \\
\end{tabular}
\caption{Model parameters, the symbols indicate: $q_0$ entering the
  definition of the rms velocity $u_0$; $N_m$ the total number
  of modes; $L_0$ and $l_{N_m-1}$ the largest and smallest model
  length scales; $\tau_0$ and $\tau_{N_m-1}$ the lrgest and the
  smallest time scales. The ratio of the simulation time step $dt$ to
  the fastest time scale $\tau_{N_m-1}$ is $dt / \tau_{N_m-1} \simeq
  60$, in both cases.}
\label{table1}
\end{table}

\subsection{Tracer pair dispersion}
\label{subsec:2p}
\begin{figure}
\centerline{\includegraphics[width=10cm]{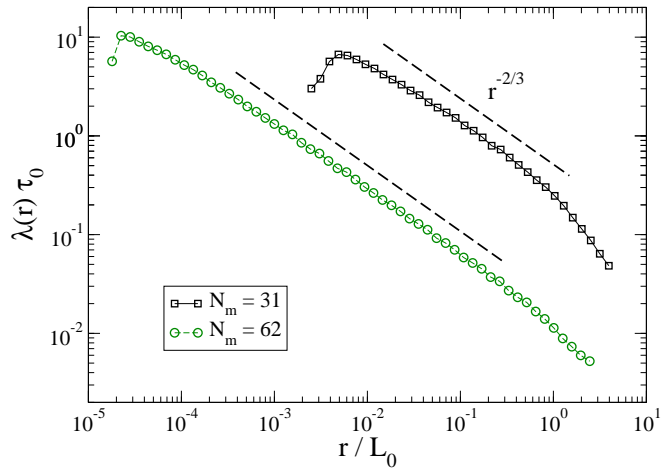}}
\caption{Log-log plot of the Finite Size Lyapunov Exponent
  $\lambda(r)$ for the numerical simulations with $N_m=31$ (black squares)
  and $N_m=62$ (green circles). The straight dashed lines indicate the
  $r^{-2/3}$ scaling regime.}
\label{fig:fsle}
\end{figure}
\begin{figure}
\centerline{\includegraphics[width=10cm]{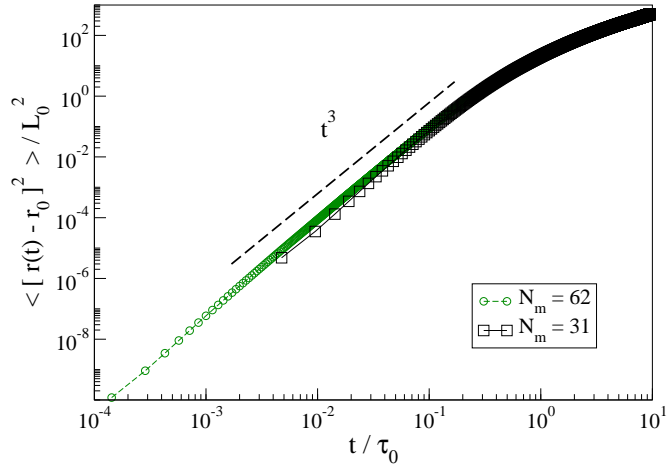}}
\caption{Log-log plot of particle relative dispersion for the
  simulations with $N_m=31$ (black squares) and $N_m=62$ (green
  circles). The straight line indicates the Richardson $t^3$ scaling
  regime for the inertial range of scales.}
\label{fig:reldisp}
\end{figure}

The separation statistics of pairs of particles, labeled $1$ and $2$,
is defined via the moments of the separation vector ${\bf r}(t)= {\bf
  x}_1(t) - {\bf x}_2(t)$. In statistically homogeneous and isotropic
turbulence, the separation distance $r = |{\bf r}|$ is the key
observable for the problem of relative dispersion. \\We first report
results on the statistics of
\begin{equation}
\langle \left[{\bf r}(t) - {\bf r}_0 \right]^2 \rangle_{r_0} = \langle \left\{ [{\bf
    x}_1(t) - {\bf x}_2(t)] - [{\bf x}_1(0) -{\bf x}_2(0)] \right\}^2
\rangle\,,
\label{eq:relative}
\end{equation}
to better highlight the scaling behaviours. Two regimes characterise the
pair dispersion for inertial-range initial distances $r_0$,
\begin{eqnarray}
&&\langle \left[{\bf r}(t) - {\bf r}_0 \right]^2 \rangle_{r_0} \simeq
t^2 S_2(r_0)\,,\\ &&\langle {\bf r}(t)^2 \rangle \simeq g\,t^3\,.
\label{eq:batchrich}
\end{eqnarray}
The first behaviour is the so-called Batchelor regime
\cite{Sawfordrev} due to the memory of the initial velocity
difference. In eqn. (\ref{eq:batchrich}), $S_2(r)$ is the Eulerian
second-order structure function at scale $r$. The Batchelor scaling
occurs on time scales of the order of the eddy-correlation time at
scale $r_0$. The second is the Richardson regime, indipendent of the
initial separation, taking place asymptotically on times scales much
larger than the eddy-turnover time at scale $r_0$, and much smaller
than the integral time scale. In Figure~\ref{fig:reldisp}, we plot
$\langle \left[{\bf r}(t) - {\bf r}_0 \right]^2 \rangle$, where the
average is taken over pairs whose initial separation $r_0$ is much
smaller that the smallest scale $l_{N_m -1}$, accordingly the
ballistic regime is very short, and the asymptotic Richardson regime
is readily observed.
The existence of Richardson regime in $3D$ turbulence is often
debated. In addition to the memory of the initial conditions, both
experimental and numerical measurements have to deal with the limited
extension of the inertial range, and with crossover regimes towards
the infra-red cut off (from inertial to large scales) and the
ultra-violet cutoff (from inertial to dissipative scales). \\
A better suited observable, allowing to partially overcome issues due
to a limited inertial range, is obtained by using fixed-scale
statistics \cite{artale}. It consists of fixing a set of thresholds,
$r_n =􏲵\rho^n r_0$, with the factor $\rho >1$ and $n=1,2,3,...$, and
then calculating the time $T$ it takes for the pair separation to
change from $r_n$ to $r_{n+1}$. If $\rho=1$ such time is also called
the {\it doubling time}. Here we compute the Finite-Size Lyapunov
Exponent (FSLE) (\cite{Cencinirev}), in terms of the mean time
$\langle T_\rho(r)\rangle $ it takes for pair separation $r$ to grow
to $\rho r$. In the present analysis, we used $\rho = 1.25$: note that
the choice of $\rho>1$ is irrelevant for the scaling properties, but
only fixes the threshold spacing and enters in the mean time
expression as a prefactor \cite{boff2d,Pof2005}. The FSLE is defined
as:
\begin{equation}
\lambda(r_n) = \frac{\log \rho}{\langle T_{\rho}(r_n) \rangle}\,.
\label{eq:FSLE}
\end{equation} 
In the limit of an infinitesimal threshold $r_n$, the FSLE recovers
the maximal Lypanuov exponent of the turbulent flow
\cite{Cencinirev}. In our model however there is no tangent space
dynamics, and so the FSLE has a well defined meaning only in the
inertial range of scales. By dimensional scaling arguments, if the
mean separation grows as $\langle r(t)^2 \rangle \sim t^3$, then the
FSLE behaves as $\lambda(r) \sim r^{-2/3}$. At scales larger than the
integral scale $L_0$, we expect $\lambda(r) \sim r^{- 2}$. \\ In
Fig. \ref{fig:fsle}, we plot the FSLE measurements for the two sets of
numerical simulations performed. Since by construction all scales $r$,
with $l_{N_m-1} < r < L_0$, belong to the inertial range, we observe
the $\lambda(r) \sim r^{-2/3}$ scaling only. At large scales, we
detect a steeper slope, associated to Taylor diffusion, where at small
scales, FSLE drops since there is no dissipative range dynamics in
this system.

\subsection{Tetrad dispersion}
\label{subsec:4p}
\begin{figure}
\centerline{\includegraphics[width=10cm]{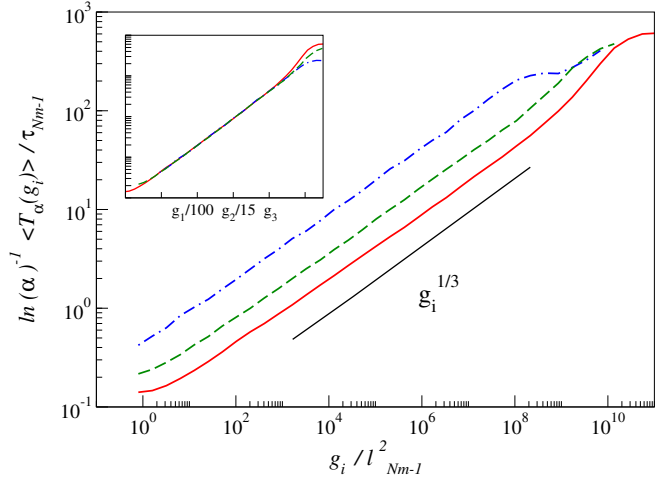}}
\caption{Mean exit-time values for the eigenvalues of the inertia
  matrix. In the inset the three curves, after rescaling $g_1$ and
  $g_2$ with the factors $100$ and $15$, respectively, to obtain an
  overlap.}
\label{fig:tetrad_et}
\end{figure}

\begin{figure}[!t]
\includegraphics[width=8cm]{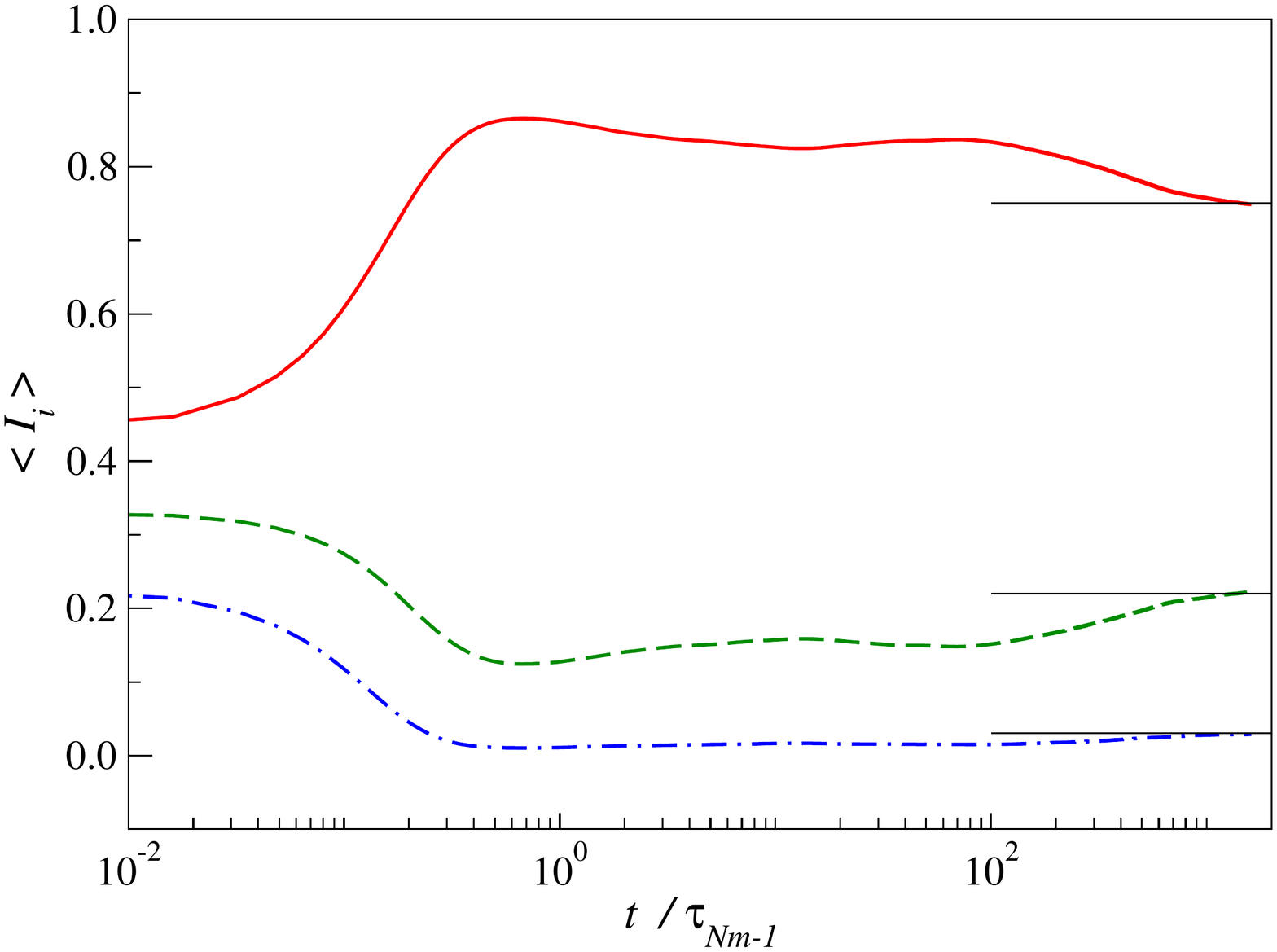}
\includegraphics[width=8cm]{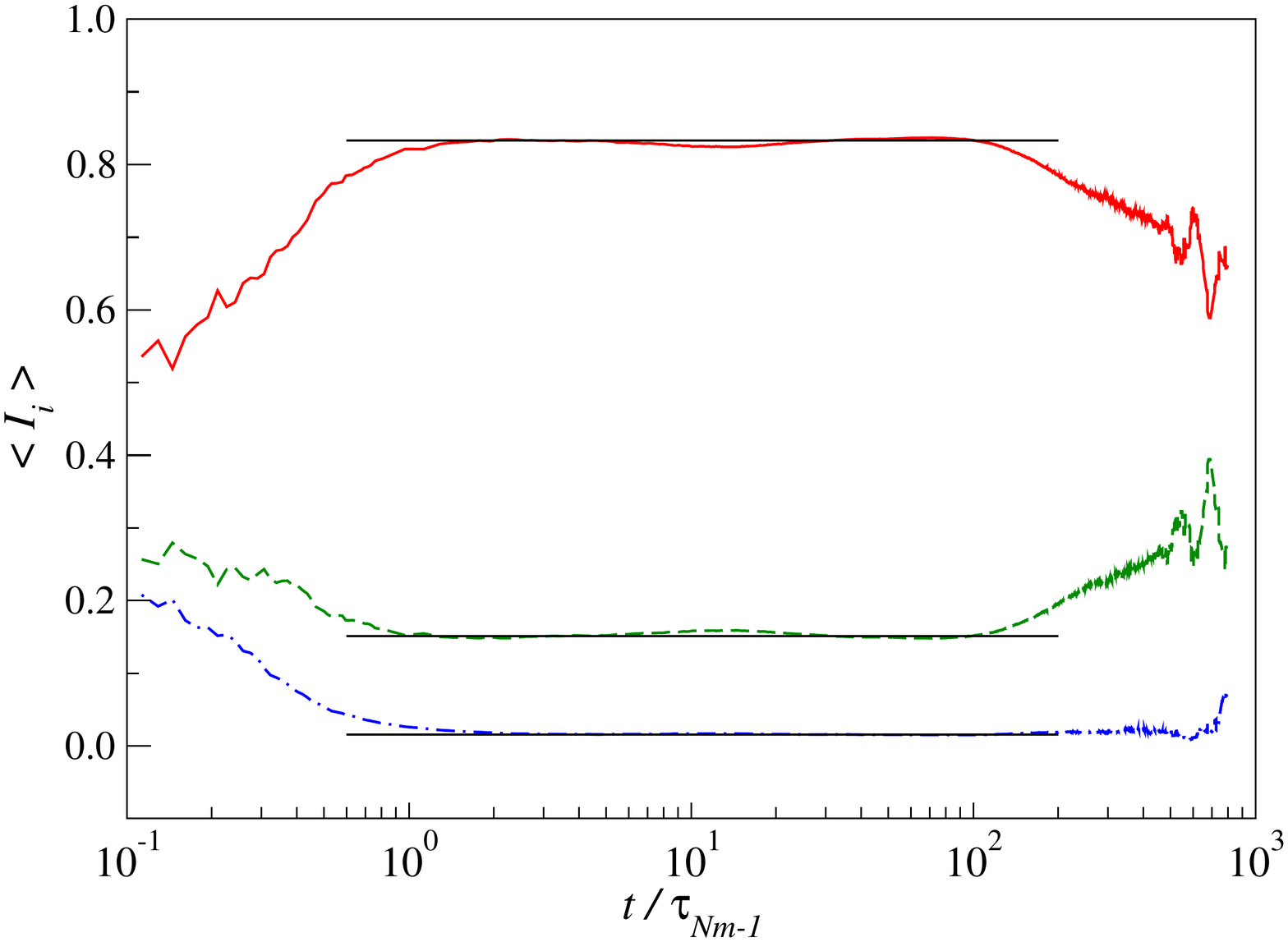}
\caption{Average shape parameters $\langle I_i \rangle$ as a function of
time. On the left: average upon all tetrads. The straight lines
indicate the asymptotic values for Gaussian distributed
particles $I_1=0.75$, $I_2=0.22$, $I_3=0.03$. On the right:
average upon tetrads whose $g_i$ belongs to the inertial range.
The straight lines are best fits for $1<t/\tau_{Nm-1}<100$: $I_1=0.833$, 
$I_2=0.151$, $I_3=0.0155$.}
\label{fig:tetrad_I}
\end{figure}

The interest in studying the displacement statistics of a bunch of
particles is that it can be used to describe moments of a passive
scalar field, satisfying an advection-diffusion equation. This has
been exploited in the past to assess the intermittent statistics of a
passive scalar field advected by a synthetic Gaussian velocity field
\cite{frismazverg}, or by a $2D$ turbulent velocity field in the
turbulent regime of inverse cascade of energy
\cite{celverg}. Unfortunately, similar results do not yet exist for
$3D$ turbulence. On the other hand, Lagrangian multi-particle motion
is very important when studying dispersion and mixing properties, both
in ideal (i.e., statistically homogeneous and isotropic) and in real
flows.\\ 
\begin{figure}
\centerline{\includegraphics[width=10cm]{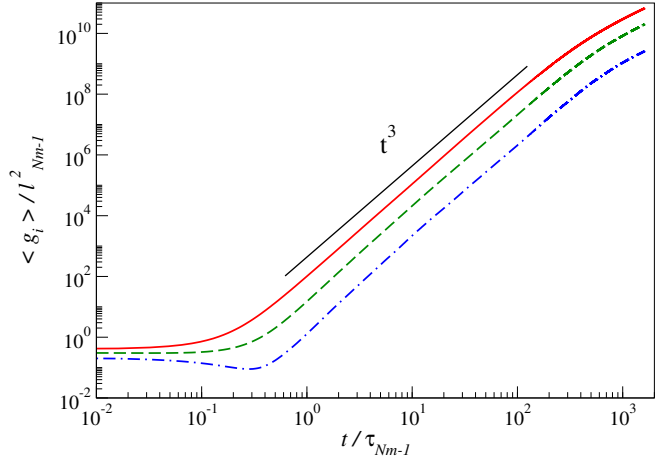}}
\caption{Log-log plot of the mean eigenvalues of the inertia matrix,
  as a function of time. The straight line indicates the $t^3$
  Richardson scaling law for the inertial range of scales.}
\label{fig:tetrad_g}
\end{figure}
Beside pair dispersion, tetrad dispersion and its modelling
have attracted much attention in the last years
\cite{pum00,Pof2005,Dev2013,Sawford2013}: indeed, the time evolution
of four tracers is the building block of a phenomenological model to
describe the Lagrangian dynamics over a volume region with
characteristic scales lying in the inertial range \cite{CPS99};
additionally it is a better candidate than a tracer pair to describe
geometrical properties in a turbulent flow, such as vortex stretching,
and vorticity/strain alignment. \\ Within the proposed model, we have
performed series of simulations with tracer particles initially
arranged on regular tetrahedron of side $\l_{N_m-1}/2$, so to have a
narrow distributions for tetrads initial conditions. Moreover, the
initial distribution of the tetrad center of mass and orientation is
random uniform in the computational domain. To achieve good
statistics, we collected $50$ simulations, each containing $100$
tetrahedron. The model parameters are those listed in Table
\ref{table1}, case $N_m=62$.\\ The evolution of the tetrad shapes in a
statistically homogeneous flow is conveniently analyzed by performing
a change of coordinate \cite{CPS99}, from the particle positions ${\bf
  x}_i$, $i=1,2,3,4$ to the reduced set of coordinates
${\boldsymbol\rho}_m$, $m=0,1,2,3$, defined as:
\begin{eqnarray}
{\boldsymbol\rho}_0 &=& \frac{1}{4}\,\sum_{i=1}^4 {\bf x}_i \,,\\
{\boldsymbol\rho}_m &=& \frac{1}{\sqrt{m(m+1)}}\,\sum_{i=1}^m ({\bf x}_{i} -{\bf x}_{m+1}) \,.
\label{eq:redcoor}
\end{eqnarray}
While the center of mass diffuses in the flow, the geometrical
information is contained in the square symmetric inertia matrix ${\bf
  I} = \boldsymbol\rho \boldsymbol\rho^T$, with column vectors
$\boldsymbol\rho_m$, $m=1,2,3$. Owing to the homogeneity of the
velocity field and of the initial tetrad distribution, the statistics
does not depend on the centre of mass ${\boldsymbol\rho}_0$.\\ The
matrix admits real positive eigenvalues, $g_i$, that can be ordered
according to: $g_1 \ge g_2 \ge g_3$.  The tetrahedron dimension is
given by $r = \sqrt{2/3\,tr({\bf I})}= \sqrt{2/3(g_1 +g_2 +g_3)}$ and
the volume is $V = 1/3 \sqrt{det({\bf I})} = 1/3 \sqrt{g_1 g_2
  g_3}$. It is convenient to introduce the adimensional quantities
$I_i = g_i/r^2$ (where $I_1+I_2+I_3=1$), whose relative values give an
indication of the tetrahedron shape. For a regular tetrahedron
$I_1=I_2=I_3=1/3$; when the four points are coplanar $I_3=0$; when
they are aligned $I_2=I_3=0$.\\ We remark that by means of a
stochastic model for tetrad dispersion, Devenish \cite{Dev2013}
recently obtained values for the $I_i$ indices in agreement with those
of Direct Numerical Simulations of $3D$ turbulence.\\ In
Fig. \ref{fig:tetrad_g} we present the temporal evolution of the mean
eigenvalues of ${\bf I}$. Numerical results show good agreement with
Richardson prediction, i.e. $\langle g_i \rangle \simeq t^3$.  This
issue is further verified by measuring fixed scale statistics. To this
aim we compute the average time $\langle T_\alpha(g_i) \rangle$ it
takes for each eigenvalue $g_i$ to increase its value of a factor
$\alpha$, with $\alpha =2$, i.e. $\langle T_\alpha(g_i) \rangle$ are
the average eigenvalues doubling times.\\ Results are plotted in
Fig. \ref{fig:tetrad_et}. They indicate the existence of a wide
inertial range, where the slope of the exit-time is $g_i^{1/3}$,
matching Richardson prediction. In addition, in the inset, the three
eigenvalues overlap after rescaling $g_1$ and $g_2$ respectively with
the factors $100$ and $15$. These scaling factors yield to $I_1 \simeq
0.862$, $I_2 \simeq 0.129$, $I_3 \simeq 0.0086$, i.e. very elongated
tetrahedron.
The existence of a range where, after rescaling on the horizontal
axis, the values of the exit-time are the same for the three
eigenvalues implies that the tetrads increase their dimension while
maintaining the same (elongated) shape. The results can be compared
with the DNS of \cite{Pof2005multi}. They show qualitative agreement,
though the values of the rescaling factors applied to achieve the
exit-times collapse are different. \\ In Fig. \ref{fig:tetrad_I}
(left) we present the behaviour of the $\langle I_i \rangle$, with
$i=1,2,3$ as a function of time. The coefficients present, over a
large time interval, values consistent with elongated tetrahedron;
then, for $t/\tau_0 \sim 1$ (i.e. $t/\tau_{N_m-1}\sim O(10^3)$) they
tend to the values obtained for tetrads formed from Gaussian
distributed particles \cite{pum00}.
In Fig. \ref{fig:tetrad_I} (right) the $\langle I_i \rangle$ are
computed selecting at each time step those $g_i$, for which the
exit-time follows Richardson scaling law, i.e.  $1< g_1/l_{Nm-1}, \,
g_2/l_{Nm-1}<10^9$, $1<g_3/l_{Nm-1}<10^8$. The figure shows the
presence of a plateau where the values of the indexes are $\langle I_1
\rangle =0.833 \pm 0.004$, $\langle I_2 \rangle =0.151 \pm 0.003$, and
$\langle I_3 \rangle =0.0155 \pm 0.0007$.  Again, there is some
discrepancy with the direct numerical simulations results for HIT,
where it was measured $\langle I_3 \rangle = 0.011 \pm 0.001$ and
$\langle I_2 \rangle = 0.135 \pm 0.003$
\cite{Pof2005}. However these values confirm the presence of elongated
structures in the inertial range, with the index $\langle I_3 \rangle$
larger with respect to the expectation value for Gaussian distributed
particles. 

\section{Conclusion}
\label{sec:conc}
A novel Lagrangian model is presented aimed at accurately reproducing
the statistical behaviour of clouds of particles dispersed in
statistically isotropic and homogeneous, incompressible turbulent
flows. The model reflects the multi-scale nature of the direct energy
cascade of $3D$ turbulence. While the model is primarily meant to be
used as a sub-grid model -it evolves fluid tracers sub-grid velocities
that are correlated according to their relative distances-, it may be
adapted to solve the Navier-Stokes equations by a Lagrangian approach
(in the spirit of smoothed-particle hydrodynamics solvers). \\ To assess the
model performances and accuracy, we presented several validations
based on comparison with recent investigation on the phenomenology of
fluid tracers in high-resolution, high-statistics Direct Numerical
Simulations. The first validation consisted in performing two
simulations that differ only by total number of modes, while the large
length and velocity scales are kept constant. We showed that the model
can reproduce Richarson law for the pair dispersion statistics. It is
important to stress that the width of the inertial range can be {\it
  apriori} fixed by tuning the sub-grid model parameters. With respect
to multi-particle statistics, we analysed the dispersion of tracers
initially located on the side points of tetrahedra. Also in this case
we could detect a good agreement with results obtained in direct
numerical simulations of homogeneous and isotropic turbulence
\cite{Pof2005multi}.\\ Two important approximations have been adopted
to build up the model: statistics is Gaussian and
self-similar. Deviations from Gaussianity could be of interest if the
tracer particle model is used to reproduce stationary statistics of
turbulent velocity increments (e.g. the four-fifth law) \cite{Pagnini}. In the
present formulation of the model, we neglected such feature and showed
that this does not affect results for pair and tetrad
dispersion. \\Neglecting intermittency may also be a limitation since
non self-similar corrections to the Richardson's picture have been
detected in the tails of pair separation distribution
\cite{BoffSok3D,SBTPRL2012}. Intermittency could be introduced by
building up synthetic multiaffine processes \cite{BBCCV98,
  toschi_bode}. This is left for future investigations.\\ Based on the
accuracy of the results, it appears that the potential of the model
for practical use is high. First of all, it can be applied, within the
restrictions discussed in the paper, to an arbitrary number of fluid
tracers and the computational cost will grow with the number of
tracers. Moreover the model parameters can be choosen to achieve the
desired extension of the inertial range. Finally, the absence of a
grid makes the method suitable also for complex situations, for
instance in the presence of free surfaces. The capabilities of the
model in more complex flows, e.g. shear and channel flows, will be a
matter of future investigations. Finally, the model may be easily
modified to describe inertial heavy point-like particles
\cite{MaxeyRiley1983}.

\begin{acknowledgments}
We acknowledge useful discussions with Luca Biferale, Ben Devenish and
Guglielmo Lacorata. I.M.M. was supported by FIRB under Grant
No. RBFR08QIP5\_001. Numerical Simulations were performed on the Linux
Cluster Socrate at CNR-ISAC (Lecce, Italy); we thank Dr. Fabio Grasso
for technical support. This work was partially supported by the
Foundation for Fundamental Research on Matter (FOM), which is part of
the Netherlands Organisation for Scientific Research (NWO).
\end{acknowledgments}

\appendix*
\section{Particle Spatial Distribution}
\begin{figure}[!t]
\centerline{\includegraphics[width=12cm]{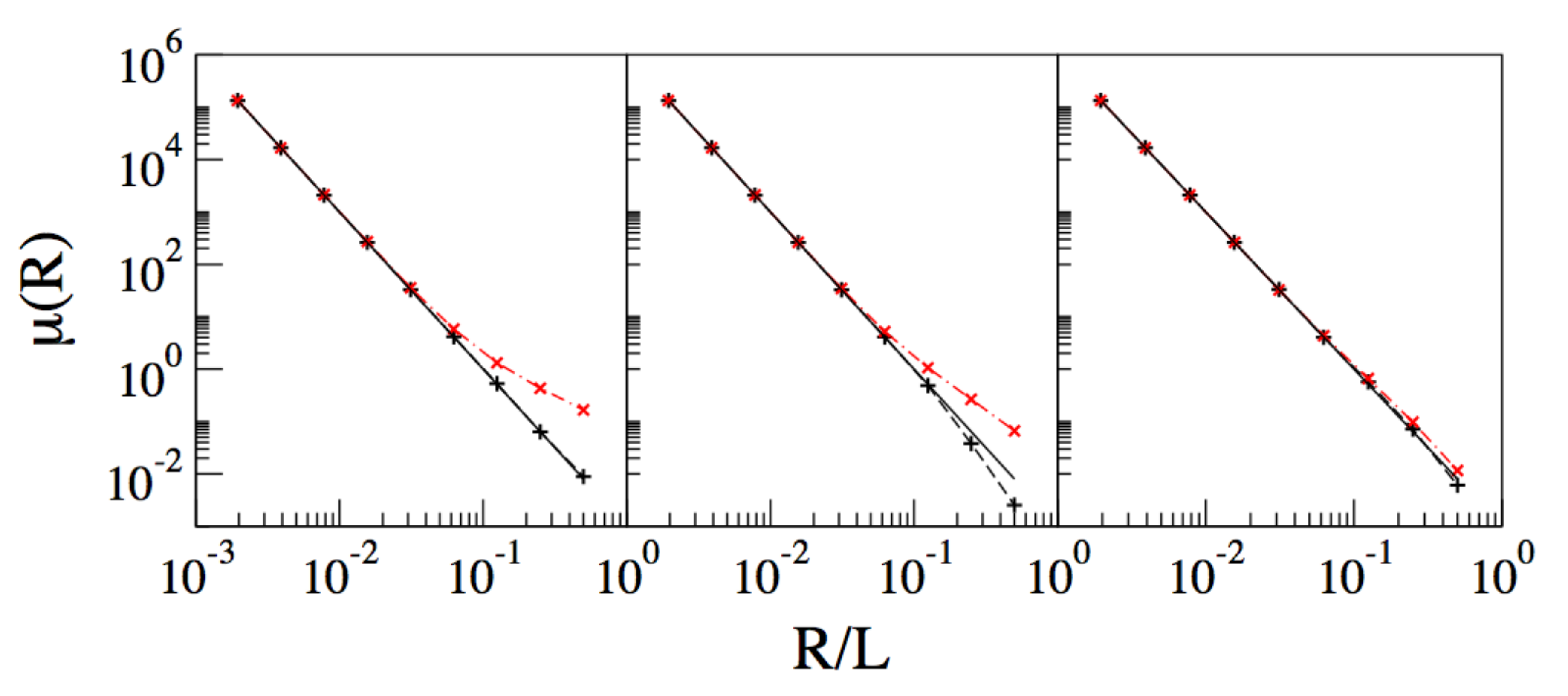}}
\caption{Coefficient $\mu(R)$ as a function of $R/L$ for simulation
with $L=2L_0$ (left); $L=4L_0$ (centre) and $L=8L_0$ (right). Results
indicate: time $t=0$ (black pluses), time $t=2\tau_0$ (red crosses) and the
uniform distribution expectation, $1/(\rho R^3)$ (black straight line).}
\label{fig:mu}
\end{figure}

In order to test particle model incompressibility, we performed the
following experiment. We seeded a periodic cubic domain with ${\cal
  N}=1000$ particles, uniformly distributed and with $N_m = 31$
velocity modes, that we followed  for a few large eddy-turnover times, $\tau_0$. An uniform
distribution of ${\cal N}$ particles in the volume ${\cal V}$ means
that, after coarse-graining the volume in cells of size $R$, the
number of particles in each cell, dubbed $n$, will be a random
variable with Poisson distribution,
\begin{equation}
p_R(n) = \frac{(\lambda_R)^n}{n!}\exp{(-\lambda_R)}\,.
\label{poisson}
\end{equation}
Here $\lambda_R = {\cal N}/(L/R)^3$ is the average number of particles
in a cell of size $R^3$ and ${\cal V} = L^3$ is the total volume
considered.  From (\ref{poisson}), it is easy to derive:
\begin{equation}
\label{stand_poisson}
\langle n^2 \rangle = \langle n \rangle^2 + \langle n \rangle. 
\end{equation}
Possible deviations from the uniform distribution can be systematically
quantified, scale by scale, in terms of the coefficient:
\begin{equation}
\mu(R) = \frac{\sigma_R^2}{\lambda_R^2} =
\frac{\langle n^2 \rangle - \langle n \rangle^2}{\langle n \rangle^2},
\label{mu}
\end{equation} 
where for a uniform distribution $\langle n \rangle = \lambda_R = \rho R^3$
where $\rho = {\cal N}/{\cal V}$ is the particle number density, and
thus $\mu(R) = 1/(\rho R^3)$. Deviation from such behaviour can be
also quantified in terms of the two-points correlation in the particle
distribution.\\ The test has been performed within three periodic
boxes of sides, respectively, $2L_0$, $4L_0$ and $8L_0$, results are
presented in Figure~\ref{fig:mu}. The plots show that, on the small
scales, uniformity is retained during the temporal evolution. However,
on the large scales, spatial correlations may develop, being more
intense when the box dimension is of the same order of the large eddy
scale, and decaying at increasing the ratio $L/L_0$. Therefore, for
$L/L_0 \gg1$, particle spread uniformly within the domain, at all
scales.

Note that, in order to take into account of molecular diffusion on the
smallest scales (i.e. scales smaller than $l_{N_m-1}$) a random
fluctuation could be added to the velocity of coincident particles.
In fact, according to our model, two or more particles, residing at
the same location at the same time, will be subjected to the same
velocity, so they will stick together for all subsequent times, while molecular diffusion would guarantee that coincident particles
always separate, see \cite{Tho90}. In practice we found that it was not needed  to include this
random diffusion because these occurance are very unlikely, as it is also indicated by
the behaviour of $\mu(R)$ on the very small scales.\\
Results of Fig.~\ref{fig:mu} have a clear interpretation: particles
tend to spread uniformly, but the velocity modulation on the large
scales induces a  spatial correlations. These can be further quantified
and controlled according to the simple arguments that
follows.\\ Starting from any initial spatial condition, when
$t>\tau_0$ the average distance of one particle to its closest
neighbour, $d_1$, can be estimated by the expression \cite{cha43}:
\begin{equation}
\label{chandra}
d_1 = \frac{1}{\pi^{1/2}}\Big[ \Gamma\Big(\frac{D}{2} + 1
  \Big)\Big]^{1/D} \Gamma \Big( 1 + \frac{1}{D} \Big) \Big(
\frac{{\cal V}}{{\cal N}} \Big)^{1/D}\,,
\end{equation}
where $\Gamma$ is the Euler Gamma function, $D=3$ the space dimension,
${\cal V}$ the volume and ${\cal N}$ the total number of particles.
Numerical results agree with the theoretical expectation for randomly
distributed particles, eq. (\ref{chandra}). The deviation of $\mu(R)$
from $1/(\rho R^{3})$ occurs at a scale $R \sim d_1$
(i.e. $R/L \sim 0.055$ for ${\cal N} =1000$), because mode velocities
on all scales larger than $d_1$ are correlated. Clearly, the larger the number of correlated modes, the stronger the deviations from
uniformity.\\ 
\begin{figure}
\centerline{\includegraphics[width=10cm]{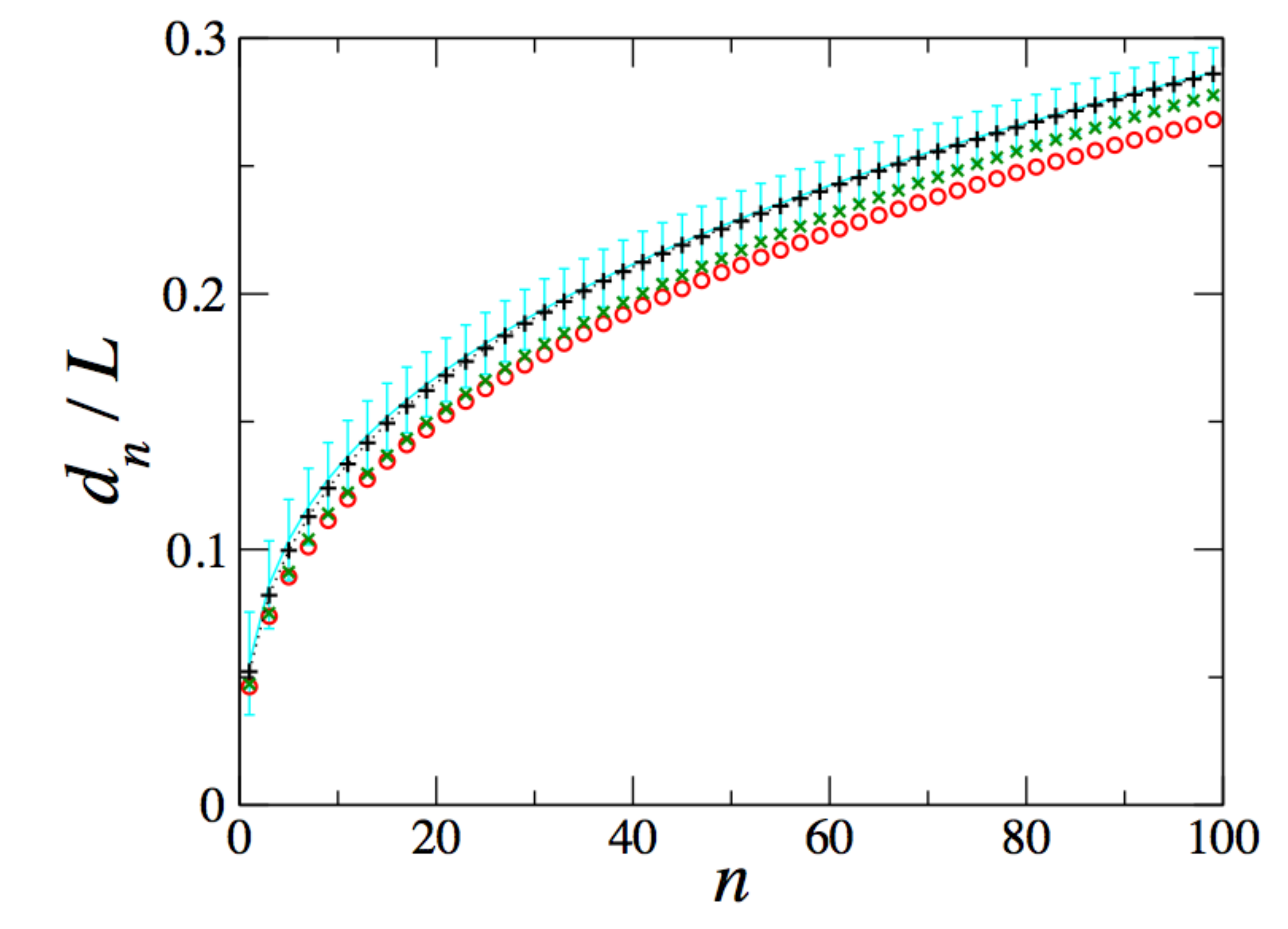}}
\caption{Average $n$-th neighbour distance for $n < 100$. The curves
  represent: the theoretical expectation with error bars (cyan),
  simulation with $L/L_0$ respectively equal to $8$ (black pluses) $4$
  (green crosses) and $2$ (red circles).}
\label{fig:dist}
\end{figure}
We remark that, when $L_0=10$ and $N_m=31$, there are
$13$ modes with length scale larger than $d_1 = 1.1$ (average particle
distance in simulations with $L=2L_0$) and only $5$ modes with length
scale larger than $d_1= 4.4$ (average distance when $L=8L_0$). This
explains the more intense deviations detected in the left plot of
Figure \ref{fig:mu} with respect to the right plot.\\ In general, the
number of correlated modes $N_c$ depends on the integral scale $L_0$,
on the model parameter $\lambda$, and on the particle density $\rho$,
according to:
\begin{eqnarray}
N_c &=& int\left(1 + log_\lambda \frac{L_0}{d_1}\right)\\ 
&=& int\left(1 + log_\lambda \left(
\frac{\sqrt{\pi} \, L_0 \, \rho^{1/3}}{\Gamma (5/2)^{1/3}
\, \Gamma (4/3)} \right)\right).
\end{eqnarray}

The average distance $d_n$ of a particle to its $n$-th neighbour can
also be computed. Recalling that one point is the $n$-th neighbor of
another one if there are exactly $n-1$ other points that are closer to
the latter than the former, the distance $d_n$, in the case of ${\cal
  N}$ uniformly distributed particles, is \cite{per96}:
\begin{eqnarray}
&&\frac{d_n}{L} = \\ &&\frac{1}{\pi^{1/2}}\Big[ \Gamma(\frac{D}{2} +1)
    \Big]^{1/D} \frac{\Gamma (n + \frac{1}{D})}{\Gamma(n)} \Big(
  \frac{1}{{\cal N}} \Big)^{1/D}\,,\nonumber
\label{eq:distn}
\end{eqnarray}
with the mean square fluctuation\,,
\begin{eqnarray}
&&\left(\frac{\Delta d_n}{L}\right)^2 = \\ 
&&\frac{1}{\pi}\left[
  \Gamma(\frac{D}{2} +1) \right]^{2/D} \left[\frac{\Gamma (n +
    \frac{2}{D})}{\Gamma(n)} - \frac{\Gamma^2(n+
    \frac{1}{D})}{\Gamma^2(n)} \right] \left( \frac{1}{{\cal N}}
\right)^{2/D}. \nonumber
\label{eq:nvar}
\end{eqnarray}
Results for the three test cases are presented in Figure
\ref{fig:dist}, as a function of the neighbour index $n$. They show
that, when ${\cal N}=1000$, only the simulation with $L/L_0=8$ has
average $n$-th neighbour distance consistent, within error bars, with
the theoretical prediction (eqs. (\ref{eq:distn}) -
(\ref{eq:nvar})). Thus, we infer that the ratio of correlated modes,
$N_c$, to the total number of modes, $N_m$, has to be rather small in
order for the model to satisfy incompressibility. The present
indication is that $N_c/N_m \simeq 1/6$ is enough to produce
uniformity at all scales (see Figure \ref{fig:mu}). When the present
model is employed as a subgrid-scale Lagrangian model this restriction
should not apply as the evolution on the larger scales will be matched
with the resolved modes of the LES.


\begin{thebibliography}{10}

\bibitem{Silv2010} H. J. S. Fernando and D. Zajic and S. Di Sabatino
  and R. Dimitrova and B. Hedquist and A. Dallman, ``Flow, turbulence
  and pollutant dispersion in urban atmospheres,'' Phys. Fluids {\bf
    22}, 051301 (2010).

\bibitem{SBTPRL2012} R. Scatamacchia and L. Biferale and F. Toschi,
  ``Extreme Events in the Dispersions of Two Neighboring Particles
  Under the Influence of Fluid Turbulence,'' Phys. Rev. Lett. {\bf
    109}, 144501 (2012).

\bibitem{toschi_bode} F. Toschi and E. Bodenschatz, ``Lagrangian
  Properties of Particles in Turbulence,'' Ann. Rev. Fluid Mech.
    {\bf 41}, 375 (2009).

\bibitem{MH2006} N.S. Holmes and L. Morawska, ``Review of Dispersion
  Modelling and its application to the dispersion of particles: An
  overview of different dispersion models available,''
  Atmos. Environ. {\bf 40}, 5902 (2006).

\bibitem{CM2011} M. Chamecki and C. Meneveau, ``Particle boundary
  layer above and downstream of an area source: scaling, simulations,
  and pollen transport,'' J. Fluid Mech. {\bf 683}, 26 (2011).

\bibitem{LMR2008} G. Lacorata and A. Mazzino and U. Rizza, ``3D
  Chaotic Model for Subgrid Turbulent Dispersion in Large Eddy
  Simulations,'' J. Atmos. Sci.  {\bf 65:7}, 2389 (2008).

\bibitem{ML2012} A. S. Lanotte and I. M. Mazzitelli, ``Scalar
  Turbulence in Convective Boundary Layers by Changing the Entrainment
  Flux,'' J. Atmos. Sci. {\bf 70}, 248 (2013).

\bibitem{berti2011} S. Berti and F. A. Dos Santos and G. Lacorata and
  A. Vulpiani, ``Lagrangian Drifter Dispersion in the Southwestern
  Atlantic Ocean,'' J. Phys. Ocean. {\bf 41}, 1659 (2011).

\bibitem{Palatella2013} L. Palatella and F. Bignami and F. Falcini and
  G. Lacorata and A. S. Lanotte and R. Santoleri, ``Lagrangian
  simulations and inter-annual variability of anchovy egg and larva
  dispersal in the in the Sicily Channel,'' accepted on
    J. Geophys. Res. - Ocea. (2014).

\bibitem{shaw2003} R. A. Shaw, ``Particle-turbulence interactions in
  atmospheric clouds,'' Ann. Rev. Fluid Mech.  {\bf 35}, 183 (2003).

\bibitem{BMB2008} M. U. Ba\"ebler and M. Morbidelli and J. Bałdyga,
  ``Modelling the breakup of solid aggregates in turbulent flows,''
  J. Fluid Mech. {\bf 612}, 261 (2008).

\bibitem{BBL2012} M. U. Ba\"ebler and L. Biferale and A. S. Lanotte,
  ``Breakup of small aggregates driven by turbulent hydrodynamical
  stress,'' Phys. Rev. E {\bf 85}, 025301(R) (2012).

\bibitem{PA2002} S. Post and J. Abraham, ``Modeling the outcome of
  drop-drop collisions in Diesel sprays,'' Int. J. of Multiphase Flow
  {\bf 28}, 997 (2002).

\bibitem{segaut} P. Segaut, {\it Large Eddy Simulation for
  Incompressible Flows} (Third ed.), Springer (2006).

\bibitem{MK2000} C. Meneveau and J. Katz, ``Scale-Invariance and
  Turbulence Models for Large-Eddy Simulation,'' Annu. Rev. Fluid
  Mech. {\bf 32(1)}, 1 (2000).

\bibitem{MoengSullivan} P.P. Sullivan and J. C. McWilliams and
  C.-H. Moeng, ``A sub-grid-scale model for large-eddy simulation of
  planetary boundary layer flows.'' Bound.-Layer Meteor. {\bf 71}, 247
  (1994).

\bibitem{ToschiLeve} E. L\'ev\^eque and F. Toschi and L. Shao and
  J.-P. Bertoglio, ``Shear-improved Smagorinsky model for large-eddy
  simulation of wall-bounded turbulent flows,'' J. Fluid Mech. {\bf
    570}, 491 (2007).

\bibitem{Ott} S. Ott and J. Mann, ``An experimental investigation of
  the relative diffusion of particle pairs in three-dimensional
  turbulent flows,'' J. Fluid Mech. {\bf 422}, 207 (2000).

\bibitem{Xunature} A. La Porta and G. A. Voth and A. M. Crawford and
  J. Alexander and E. Bodenschatz, ``Fluid particle accelerations in
  fully developed turbulence,'' Nature {\bf 409}, 1017 (2001).

\bibitem{PintonPRL} N. Mordant and P. Metz and O. Michel and
  J. F. Pinton, ``Measurement of Lagrangian velocity in fully
  developed turbulence,'' Phys. Rev. Lett. {\bf 87}, 214501 (2001).

\bibitem{IK2002} T. Ishihara and Y. Kaneda, ``Relative diffusion of a
  pair of fluid particles in the inertial subrange of turbulence,''
  Phys. Fluids {\bf 14(11)}, L69 (2002).

\bibitem{YeBo2004} P. K. Yeung and M.S. Borgas, ``Relative dispersion
  in isotropic turbulence: part 1. Direct numerical simulations and
  Reynolds number dependence,'' J. Fluid Mech. {\bf 503}, 125 (2004).

\bibitem{BoffSok3D} G. Boffetta and I. M. Sokolov, ``Relative
  dispersion in fully developed turbulence: The Richardson’s law and
  intermittency corrections,'' Phys. Rev. Lett. {\bf 88}, 094501
  (2002).

\bibitem{PRL2004} L. Biferale and G. Boffetta and A. Celani and
  B.J. Devenish and A. Lanotte and F. Toschi, ``Multifractal
  statistics of Lagrangian velocity and acceleration in turbulence,''
  Phys. Rev. Lett. {\bf 93}, 064502 (2004).

\bibitem{Pof2005} L. Biferale and G. Boffetta and A. Celani and
  B.J. Devenish and A. Lanotte and F. Toschi, ``Lagrangian statistics
  of particle pairs in homogeneous isotropic turbulence,''
  Phys. Fluids {\bf 17}, 115101 (2005).

\bibitem{Yeungrev} P.K. Yeung, ``Lagrangian investigations of
  turbulence,'' Annu. Rev. Fluid Mech. {\bf 34}, 115 (2002).

\bibitem{Sawfordrev} B. Sawford, ``Turbulent relative dispersion,''
  Annu. Rev. Fluid Mech. {\bf 33}, 289 (2001).

\bibitem{Collinsrev} J.P.L.C. Salazar and L.R. Collins, ``Two-Particle
  Dispersion in Isotropic Turbulent Flows,'' Annu. Rev. Fluid
  Mech. {\bf 41}, 435 (2009).

\bibitem{JFM2010} J. Bec and L. Biferale and A.S. Lanotte and
  A. Scagliarini and F. Toschi, ``Turbulent pair dispersion of
  inertial particles,'' J. Fluid Mech. {\bf 645}, 1 (2010).

\bibitem{Weil2004} J. C. Weil and P.P. Sullivan and C.-H. Moeng, ``The
  Use of Large-Eddy Simulations in Lagrangian Particle Dispersion
  Models,'' J. Atmos. Sci. {\bf 61}, 2877 (2004).

\bibitem{JH2013} G. Jin and G.-W. He, ``A nonlinear model for the
  subgrid timescale experienced by heavy particles in large eddy
  simulation of isotropic turbulence with a stochastic differential
  equation,'' New J. Phys. {\bf 15}, 035011 (2013).
  
\bibitem{FGVrev} G. Falkovich and K. Gaw\c{e}dzki and M. Vergassola,
  ``Particles and fields in fluid turbulence,'' Rev. Mod. Phys. {\bf
  73}, 913 (2001).

\bibitem{CPS99} M. Chertkov and A. Pumir and B. Shraiman, ``Lagrangian
  tetrad dynamics and the phenomenology of turbulence,'' Phys. Fluids
  {\bf 11}, 2394 (1999).

\bibitem{Pof2005multi} L. Biferale and G. Boffetta and A. Celani and
  B.J. Devenish and A. Lanotte and F. Toschi, ``Multiparticle
  dispersion in fully developed turbulence,'' Phys. Fluids {\bf 17},
  111701 (2005).

\bibitem{Xu2007} H. Xu and N. T. Ouellette and E. Bodenschatz,
  ``Evolution of geometric structures in intense turbulence,'' New
  J. Phys. {\bf 10}, 013012 (2008).

\bibitem{HYS2011} J. F. Hackl and P.K. Yeung and B.L. Sawford,
  ``Multi-particle and tetrad statistics in numerical simulations of
  turbulent relative dispersion,'' Phys. Fluids {\bf 23}, 065103
  (2011).
  
\bibitem{PRLall} A. Arn\'eodo {\it et al.}, ``Universal Intermittent
  Properties of Particle Trajectories in Highly Turbulent Flows,''
  Phys. Rev. Lett. {\bf 100}, 254504 (2008).

\bibitem{Tho87} D. J. Thomson, ``Criteria for the selection of
  stochastic models of particle trajectories in turbulent flows,''
  J. Fluid Mech. {\bf 180}, 529 (1987).

\bibitem{KD88} H. Kaplan and N. Dinar, ``A three dimensional
  stochastic model for concentration fluctuation statistics in
  isotropic homogeneous turbulence,'' J. Comput. Phys. {\bf 79}, 317
  (1998).

\bibitem{Tho90} D. J. Thomson, ``A stochastic model for the motion of
  particle pairs in isotropic high-Reynolds number turbulence, and its
  application to the problem of concentration variance,'' J. Fluid
  Mech. {\bf 210}, 113 (1990).

\bibitem{K97} O. A. Kurbanmuradov, ``Stochastic Lagrangian models for
  two-particle relative dispersion in high-Reynolds number
  turbulence,'' Monte Carlo Meth. Appl. {\bf 3}, 37 (1997).
  
\bibitem{DevTho2013} B. J. Devenish and D. J. Thomson, ``A Lagrangian
  stochastic model for tetrad dispersion'', Journ. Turb. {\bf 14},3
  (2013).

\bibitem{Dev2013} B. J. Devenish, ``Geometrical Properties of
  Turbulent Dispersion,'' Phys. Rev. Lett. {\bf 110}, 064504 (2013).
  
\bibitem{Fung1992} J. C. H. Fung and J. C. R. Hunt and N. A. Malik and
  R. J. Perkins, ``Kinematic simulation of homogeneous turbulent flows
  generated by unsteady random Fourier modes,'' J. Fluid Mech. {\bf
    236}, 281 (1992).

\bibitem{FV1998} J. C. H. Fung and J. C. Vassilicos, ``Two-particle
  dispersion in turbulentlike flows,'' Phys. Rev. E {\bf 57}, 1677
  (1998).

\bibitem{Chaves} M. Chaves and K. Gaw\c{e}dzki and P. Horvai and
  A. Kupiainen and M. Vergassola, ``Lagrangian dispersion in Gaussian
  self-similar velocity ensembles,'' J. Stat. Phys. {\bf 133}, 643
  (2004).

\bibitem{DevThom2005} D. J. Thomson and B. J. Devenish, ``Particle
  pair separation in kinematic simulations,'' J. Fluid Mech. {\bf
    526}, 277 (2005).

\bibitem{Eyinksweep} G. L. Eyink and D. Benveniste, ``Suppression of
  particle dispersion by sweeping effects in synthetic turbulence,''
  Phys. Rev. E {\bf 87(2)}, 023011 (2013).

\bibitem{K2006} J. G. M. Kuerten, ``Subgrid modeling in particle-laden
  channel flow,'' Phys. Fluids {\bf 18}, 025108 (2006).

\bibitem{soldati_salvetti} C. Marchioli and M.V. Salvetti and
  A. Soldati, ``Some issues concerning Large-Eddy Simulation of
  inertial particle dispersion in turbulent bounded flows,''
  Phys. Fluids {\bf 20}, 040603 (2008).

\bibitem{toschi_donini} E. Calzavarini and A. Donini and V. Lavezzo
  and C. Marchioli and E. Pitton and A. Soldati and F. Toschi, ``On
  the Error Estimate in Sub-Grid Models for Particles in Turbulent
  Flows,'' in {\it Direct and Large-Eddy Simulation VIII. ERCOFTAC
    Series}, Springer Netherlands, {\bf 15}, 171--176 (2011).

\bibitem{Mon2012} J. J. Monaghan, ``Smoothed Particle Hydrodynamics
  and Its Diverse Applications,'' Annu. Rev. Fluid Mech. {\bf 44}, 323
  (2012).

\bibitem{Sawford2013} B. L. Sawford and S. B. Pope and P. K. Yeung,
  ``Gaussian Lagrangian stochastic models for multi-particle
  dispersion,'' Phys. Fluids {\bf 25}, 055101 (2013).

\bibitem{bor94} M. S. Borgas and B. L. Sawford, ``A family of
  stochastic models for two-particle dispersion in isotropic
  homogeneous stationary turbulence,'' J. Fluid Mech. {\bf 279}, 69 (1994).

\bibitem{Burgener2012} T. Burgener and D. Kadau and H. Herrmann,
  ``Particle and particle pair dispersion in turbulence modeled with
  spatially and temporally correlated stochastic processes,''
  Phys. Rev. E {\bf 86}, 046308 (2012).

\bibitem{frisch} U. Frisch, {\it Turbulence. The legacy of
  A. N. Kolmogorov}, Cambridge University Press, New York (1995).

\bibitem{gil96} D. T. Gillespie, ``Exact numerical simulation of the
  Ornstein-Uhlenbeck process and its integral,'' Phys. Rev. E {\bf
    54}, 2084 (1996).

\bibitem{pope00} S. B. Pope, {\it Turbulent Flows}, Cambridge
  University Press (2000).
         
\bibitem{Men1996} C. Meneveau, ``Transition between viscous and
  inertial-range scaling of turbulence structure functions,''
  Phys. Rev. E {\bf 54}, 3657 (1996).

\bibitem{Artale} V. Artale and G. Boffetta and A. Celani and
  M. Cencini and A. Vulpiani, ``Dispersion of passive tracers in
  closed basins: Beyond the diffusion coefficient,'' Phys. Fluids {\bf
    9}, 3162 (1997).

\bibitem{Cencinirev} M. Cencini and A. Vulpiani, ``Finite Size
  Lyapunov Exponent: review on applications,'' J. Phys. A:
  Math. Theor. {\bf 46}, 254019 (2013).

\bibitem{boff2d} G. Boffetta and I. M. Sokolov, ``Statistics of
  two-particle dispersion in two-dimensional turbulence,''
  Phys. Fluids {\bf 14}, 3224 (2002).
  
\bibitem{frismazverg} U. Frisch and A. Mazzino and M. Vergassola,
  ``Intermittency in Passive Scalar Advection,'' Phys. Rev. Lett. {\bf
  80}, 5532 (1998).
  
\bibitem{celverg} A. Celani and M. Vergassola, ``Statistical geometry
  in scalar turbulence,'' Phys. Rev. Lett.q {\bf 86(3)}, 424 (2001).
  
\bibitem{pum00} A. Pumir and B. I. Shraiman and M. Chertkov,
  ``Geometry of Lagrangian dispersion in turbulence,''
  Phys. Rev. Lett. {\bf 85}, 5324 (2000).

\bibitem{Pagnini} G. Pagnini, ``Lagrangian stochastic models for
  turbulent relative dispersion based on particle pair rotation,''
  J. Fluid Mech. {\bf 616}, 357 (2008).

\bibitem{BBCCV98} L. Biferale and G. Boffetta and A. Celani and
  A. Crisanti and A. Vulpiani, ``Mimicking a turbulent signal:
  Sequential multiaffine processes,'' Phys. Rev. E {\bf 57(6)}, R6261
  (1998).

\bibitem{MaxeyRiley1983} M. R. Maxey and J. Riley, ``Equation of
  motion of a small rigid sphere in a nonuniform flow,'' Phys. Fluids
  {\bf 26}, 883 (1983).

\bibitem{cha43} S. Chandrasekar, Rev. Mod. Phys. {\bf 15}, 1 (1943).

\bibitem{per96} A. G. Percus and O. C. Martin, ``Finite Size and
  Dimensional Dependence in the Euclidean Traveling Salesman
  Problem,'' Phys. Rev. Lett. {\bf 76}, 1188 (1996).
\end{thebibliography}
\end{document}